%% file: reporting.tex
\documentclass[letterpaper,twocolumn,10pt]{article}
\usepackage{usenix2023_SOUPS}

\usepackage{tikz}
\usepackage{amsmath}
\usepackage{graphicx}
\usepackage{soul}
\usepackage{caption}
\usepackage{subcaption}
\usepackage{multirow}
\usepackage{enumitem}
\usepackage{booktabs}

\newcommand{\leijie}[1]{#1}

\newcommand{\partiNum}{16}
\begin{document}

\date{}


\title{\Large{\textbf{"Is Reporting Worth the Sacrifice of Revealing What I’ve Sent?": Privacy Considerations When Reporting on End-to-End Encrypted Platforms}}}

\def\plainauthor{Author name(s) for PDF metadata. Don't forget to anonymize for submission!}

\author{
{\rm Anonymous Author(s)}
}

\author{
{\rm Leijie Wang}\\
University of Washington
\and
{\rm Ruotong Wang}\\
University of Washington
\and
{\rm Sterling Williams-Ceci}\\
Cornell University
\and
{\rm Sanketh Menda}\\
Cornell Tech
\and
{\rm Amy X. Zhang}\\
University of Washington
} 

\maketitle

\thecopyright

\input{sections/00_abstract.tex}

\input{sections/01_introduction.tex}


\input{sections/02_related_work.tex}


\input{sections/03_methodology.tex}

\input{sections/04_findings.tex}

\input{sections/05_discussion.tex}

\input{sections/06_limitations.tex}

\input{sections/07_conclusion.tex}
\input{sections/08_acknowledge}

\bibliographystyle{plain}
\bibliography{reporting}

\input{sections/09_appendix.tex}

\end{document}

%% file: sections/00_abstract.tex
\begin{abstract}
User reporting is an essential component of content moderation on many online platforms---in particular, on end-to-end encrypted (E2EE) messaging platforms where platform operators cannot proactively inspect message contents. However, users' privacy concerns when considering reporting may impede the effectiveness of this strategy in regulating online harassment. In this paper, we conduct interviews with 16 users of E2EE platforms to understand users' mental models of how reporting works and their resultant privacy concerns and considerations surrounding reporting. 
We find that users expect platforms to store rich longitudinal reporting datasets, recognizing both their promise for better abuse mitigation and the privacy risk that platforms may exploit or fail to protect them. We also find that users have preconceptions about the respective capabilities and risks of moderators at the platform  versus community level---for instance, users trust platform moderators more to not abuse their power but think community moderators have more time to attend to reports. These considerations, along with perceived effectiveness of reporting and how to provide sufficient evidence while maintaining privacy, shape how users decide whether, to whom, and how much to report. We conclude with design implications for a more privacy-preserving reporting system on E2EE messaging platforms.

\end{abstract}

%% file: sections/01_introduction.tex
\section{Introduction}





The emerging threats of online harassment and other offensive behaviors pose significant challenges to online platforms. A 2021 Pew survey found that 41\% of Americans reported personally experiencing harassment and bullying online~\cite{pewsurvey}. Despite the deployment of algorithms by online platforms (e.g., Facebook~\cite{FacebookAlgorithm}, Reddit~\cite{Chandrasekharan2019}) to detect abusive messages proactively, user reporting remains a widely used strategy across platforms to tackle online harassment~\cite{crawford2016flag}. After users report abusive messages, human moderators review reports and decide whether to sanction the reported user.
 
Compared to other online platforms, end-to-end encrypted (E2EE) messaging platforms such as WhatsApp must rely more heavily on user reporting to regulate online harassment. As E2EE prevents third parties from accessing conversations without users' consent, platforms cannot deploy algorithms to detect abusive messages proactively unless they use client-side scanning, an approach that violates privacy guarantees~\cite{kamara2022outside} and creates new privacy risks for users~\cite{abelson2021bugs}. Hence, user reporting is considered the most privacy-preserving moderation approach for E2EE platforms~\cite{kamara2022outside, pfefferkorn2022content}. 

However, while reporting can be used to safeguard a user's privacy in the face of abuse, it also carries privacy risks of its own.
On the one hand, reporting helps to protect users' privacy when abusers expose their sensitive information (e.g., intimate photos or sexual orientation) to a broader audience~\cite{snyder2017fifteen, lenhart2016nonconsensual}. In such cases, reporting these abusive messages so that they get removed can help prevent further dissemination of users' personal information~\cite{van2016help}. 
On the other hand, user reporting also poses new privacy risks---while reporting does not violate E2EE privacy guarantees as it is user-initiated~\cite{kamara2022outside}, it may expose private information to platforms and moderators. For instance, if users believe the context around the reported message is shared with moderators, they may hesitate to report if sensitive personal information is exposed within the context~\cite{abu2017obstacles}. 
\leijie{
Such situations might arise more frequently in the context of online harassment where the harasser is known to the user, such as an ex-partner or family member~\cite{pewsurveyprivacy,mahar2018squadbox}.}
Similarly, journalists might be concerned about exposing metadata such as account and device information in a report, which could disclose the identities of their sources to E2EE platforms, even if this information is legitimately useful for making informed moderation decisions~\cite{lerner2017confidante, mcgregor2015investigating}.

In this paper, we seek to understand people's privacy concerns when considering reporting on end-to-end encrypted (E2EE) messaging platforms. Inspired by prior research that suggests people's mental models of technologies influence their privacy behaviors~\cite{de2016expert, krombholz2019if}, we start by investigating people's mental models of user reporting, including their assumptions regarding \textit{data flows}, or what data is shared with E2EE platforms and moderators, and how the data is stored and used by platforms. In particular, we are interested in the following two research questions.\\[2mm]
\begin{tabular}{lp{7cm}}
    RQ1 & What are users’ mental models of reporting unwanted messages on E2EE messaging platforms?\\
    RQ2 & What privacy concerns and considerations do users have when they make reporting decisions on E2EE messaging platforms?
\end{tabular}

We conducted \partiNum{} semi-structured interviews with users of E2EE platforms. To help users articulate their mental models of reporting, we provided participants with cards labeled with stakeholders (e.g., platform moderators, community moderators), data (e.g., reported message, account information), and moderation actions (e.g., delete messages, ban account). We then invited them to organize these cards on a digital board to illustrate a reporting procedure while speaking aloud. As users' privacy considerations are often grounded in their reporting decisions, we created a series of hypothetical scenarios involving abusive messages that also expose different kinds of personal information (Fig. \ref{harassment scenarios}) and asked participants to talk through their reporting decisions in these scenarios.

We find that participants assume platforms already collect account and device information and that many platforms also store longitudinal report history and some context for reported messages. While participants believe that platforms will use data from user reports to build more sophisticated anti-harassment tools, they also worry that platforms may misuse this data in ways that benefit the platform at the cost of users' privacy. We also observe that participants have assumptions about the respective strengths and risks of platform moderators versus community moderators. Platform moderators are assumed to be more distant and professional, and are thus more trusted with private information. Meanwhile, community moderators are assumed to be more familiar and therefore present greater privacy risks, even as they may also have more context and more time to properly address reports.

Finally, we find participants make nuanced decisions about whether, to whom, and how much to report based on trade-offs between privacy risks and protections. 
When reporting, participants want to share just enough information for moderators to make informed decisions, though they are willing to share more if the report is anonymized or they trust platforms. But sometimes reporting is perceived as too much of a sacrifice in terms of privacy for too little gain. Based on these findings, we argue that a more privacy-preserving reporting system on E2EE platforms should provide more granularity for users to tailor their reports, enable more flexible interaction between stakeholders to ensure informed procedures, and have more transparency to cultivate users' mental models of reporting.

%% file: sections/02_related_work.tex
\section{Related Work}
\subsection{Combating online harassment and hate with user reporting}
Considerable research has established the pervasiveness of online harassment and hate for users of social media, particularly those from marginalized communities~\cite{pewsurvey, thomas2021sok, ashktorab2016designing}. A Pew survey in 2017 found that 41\% of Americans reported personally experiencing varying degrees of harassment and bullying online \cite{pewsurvey}. These abusive behaviors range from trolling that intentionally provokes audiences with inflammatory remarks~\cite{cheng2017anyone} to ``SWATing,'' in which attackers falsely report emergencies to send police to the target's address~\cite{SWATing}.

User reporting is an essential defense against online harassment and hate~\cite{thomas2021sok}.
In this work, we distinguish between reporting to platform moderators versus to community moderators. Here, \textit{communities} refer to groups of multiple chat rooms (e.g., a Matrix community~\cite{MatrixCommunity} or a WhatsApp community~\cite{WhatsAppCommunity}, or in the non-E2EE setting, a Slack workspace~\cite{SlackWorkspace} or a Discord server~\cite{DiscordServer}). At the platform level, nearly all platforms maintain reporting systems that enable users to send unwanted messages to platform moderators, who are employed to review user reports and make platform-wide moderation decisions according to platform policies regarding impermissible content~\cite{roberts2016commercial, FacebookLabor}. At the community level, each community can establish ad-hoc ways to receive user reports. For example, on Discord, users may report to community moderators via direct messages, dedicated channels, or emails~\cite{reportDiscord}. Community moderators are often community members elected or appointed to make community-specific moderation actions in accordance with community guidelines about (un)favorable behaviors~\cite{lampe2004slash, blackburn2014stfu}.
Crawford and Gillespie argue that user reporting represents interactions between users, platforms, algorithms, and broader political forces~\cite{crawford2016flag}. In our project, we delve deeper into these interactions by exploring users' perceptions of how data flows from users to communities, platforms, and moderators in reporting systems.

Different platforms have implemented different forms of reporting procedures and data records. For instance, a recent survey found that most online platforms utilize both account information (e.g., email address, account username, and the frequency of account actions) and device information (e.g., IP address) for content moderation~\cite{pfefferkorn2022content}. In the crowdsourced moderation system of League of Legends, moderators have access to the entire chat log during the match but players' handles and social contacts are removed to protect privacy~\cite{blackburn2014stfu}. 
Similarly, on Reddit, the identity of the reporter is kept anonymous to community moderators but known to platform admins~\cite{gilbert2023towards}.
In contrast, moderators on E2EE platforms have more restricted access to chats. For example, the content of the reported message is not disclosed to moderators on Signal and Matrix~\cite{MatrixReport}. On WhatsApp, reporting an account forwards the last five messages from one's conversation with them to moderators~\cite{WhatsAppReport}, while this number is 30 for Facebook Messenger~\cite{fbmessenger}.
However, reporting systems have overall been criticized for being opaque~\cite{kou2021flag}. It is unclear how much of the public information about reporting systems is known to users or what they imagine happens when they submit a report; these questions form the starting point for our study.

While it may be more privacy-preserving to limit the amount of information shared in a user report, additional context can be important for moderators to determine a course of action. Indeed, the user reporting system itself can be co-opted to further abuse~\cite{matias2015reporting}.
For instance, some platforms hide content that has received many reports until a moderator can review it, and moderators may also be convinced to take content down if enough users have reported it -- this can motivate groups to silence others by mass reporting content they dislike~\cite{zhao2023let}. Bad-faith reporters can also try to distort information in their reports.
Finally, reports can be used to abuse moderators or waste their time. For instance, community moderators on Reddit receive anonymous reports, opening them to harassment with little risk of sanction~\cite{gilbert2023towards}.
We grapple with the trade-offs in preserving user privacy or allowing users to customize what they share in a report in our Discussion \ref{design implications discussion}.

\subsection{User reporting on E2EE platforms}
E2EE messaging platforms like WhatsApp, Signal, or iMessage are popular among people for private communication. Much like non-E2EE social platforms, online harassment is also a problem on E2EE platforms. A 2022 survey by the ADL found that 12\% of adults and 15\% of teens have experienced harassment on WhatsApp~\cite{adlsurvey}, one of the most common E2EE messaging apps. E2EE platforms rely more on user reporting to regulate online harassment. Despite the increasing use of algorithms to proactively detect abusive messages across non-E2EE platforms~\cite{FacebookAlgorithm, YoutubeAlgorithm, Chandrasekharan2019}, the lack of access to messages in E2EE conversations without users' consent makes algorithmic detection impossible~\cite{Childwelfare}. Thus, user reporting in E2EE settings is considered a crucial moderation approach to keep platforms alerted to abuse while still preserving privacy and security ~\cite{kamara2022outside, pfefferkorn2022content}. To enable user reporting on E2EE platforms, cryptographic protocols such as message franking~\cite{facebookwhitepaper, dodis2018fast} allow platform moderators to verify that the sender sent the reported message while also providing deniability to entities other than moderators~\cite{tyagi2019asymmetric}.

\vspace{-3mm}
\subsubsection{Privacy risks of user reporting on E2EE platforms}
However, it turns out that 47\% of people who have experienced harassment do not bother to report it~\cite{adlsurvey}. While prior research indicates perceived ineffectiveness as one of the primary reasons why people fail to report abuse~\cite{kou2021flag}, in this work, we highlight privacy as another important but often neglected factor, especially for users on E2EE platforms who are more invested in maintaining privacy~\cite{kamara2022outside, pfefferkorn2022content}.

First, people may be reluctant to report due to private information revealed in the course of a conversation where the harassment occurred. For instance, the context around the reported message might include personally identifiable information or political views that people are unwilling to disclose to third parties~\cite{abu2017obstacles, ackerman1999privacy}. These situations arise more frequently when a harasser is known to the recipient. A Pew survey found that nearly half of Americans (46\%) who have experienced online harassment say they know the harasser, including acquaintances (26\%), family members (11\%), and ex-romantic partners (7\%)~\cite{pewsurveyprivacy}. Indeed, E2EE messaging platforms are often used for conversations among people who know each other and for sharing sensitive information.

In addition to privacy concerns around platform access to private information that can get leaked, sold, or shared, users may also need to consider the privacy risks of moderators as attackers~\cite{thomas2021sok}. With privileged access to private information in reports, moderators could carry out attacks like ``doxxing'' where targets' personal information (e.g., sexual identity and intimate photos) is exposed to a broader audience~\cite{snyder2017fifteen, lenhart2016nonconsensual, sinclair2022sa}, or surveillance where targets' devices or accounts are compromised for monitoring purposes~\cite{farinholt2017catch}. Luca et al. observed that users on E2EE platforms are more worried about the leakage of their sensitive information to people they know than to unknown entities~\cite{de2016expert}. As users have more interactions with community moderators than platform moderators, they may have more privacy concerns about sharing information with community moderators via reporting.

\subsubsection{Mental models of user reporting}
Prior work has suggested how users' mental models of technologies may influence their privacy behaviors and concerns~\cite{de2016expert, krombholz2019if}. In this work, we also observe users  struggling to understand what data is shared and how it is stored and used during the reporting process. This may be due to the degree to which online platforms maintain opaque data policies about their reporting systems~\cite{kou2021flag}. In addition, users are shielded from the decision-making process, with little insight into how moderators use information shared to reach a decision, or even whether they actually make a decision~\cite{crawford2016flag}. 

Further, a limited understanding of E2EE potentially complicates users' mental models of how reporting works on E2EE platforms. Previous studies have shown that users lack confidence and accuracy in their mental models of E2EE platforms~\cite{schroder2016signal, abu2017security}. For example, Abu-Salma et al. found that a considerable number of users believe that landline phone calls are not less secure than E2EE communications~\cite{abu2018exploring} and that their E2EE communications are vulnerable to eavesdropping by determined attackers~\cite{gerber2018finally, naiakshina2016poster}.

\begin{figure*}\scriptsize
    \centering
    \begin{subfigure}[t]{0.45\textwidth}
        \centering
        \includegraphics[width=\textwidth]{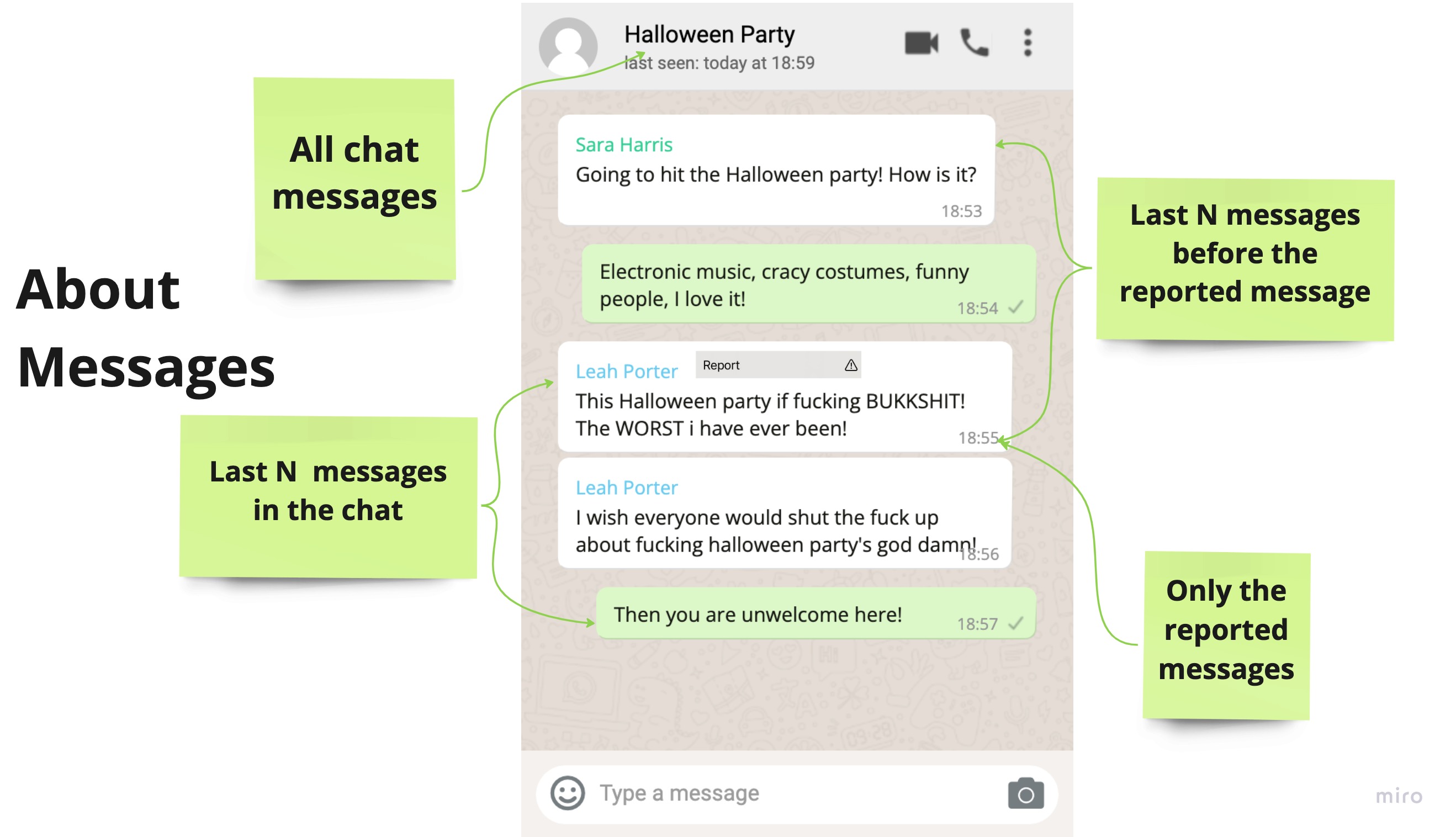}
    \end{subfigure}
    \begin{subfigure}[t]{0.45\textwidth}
        \centering
        \includegraphics[width=\textwidth]{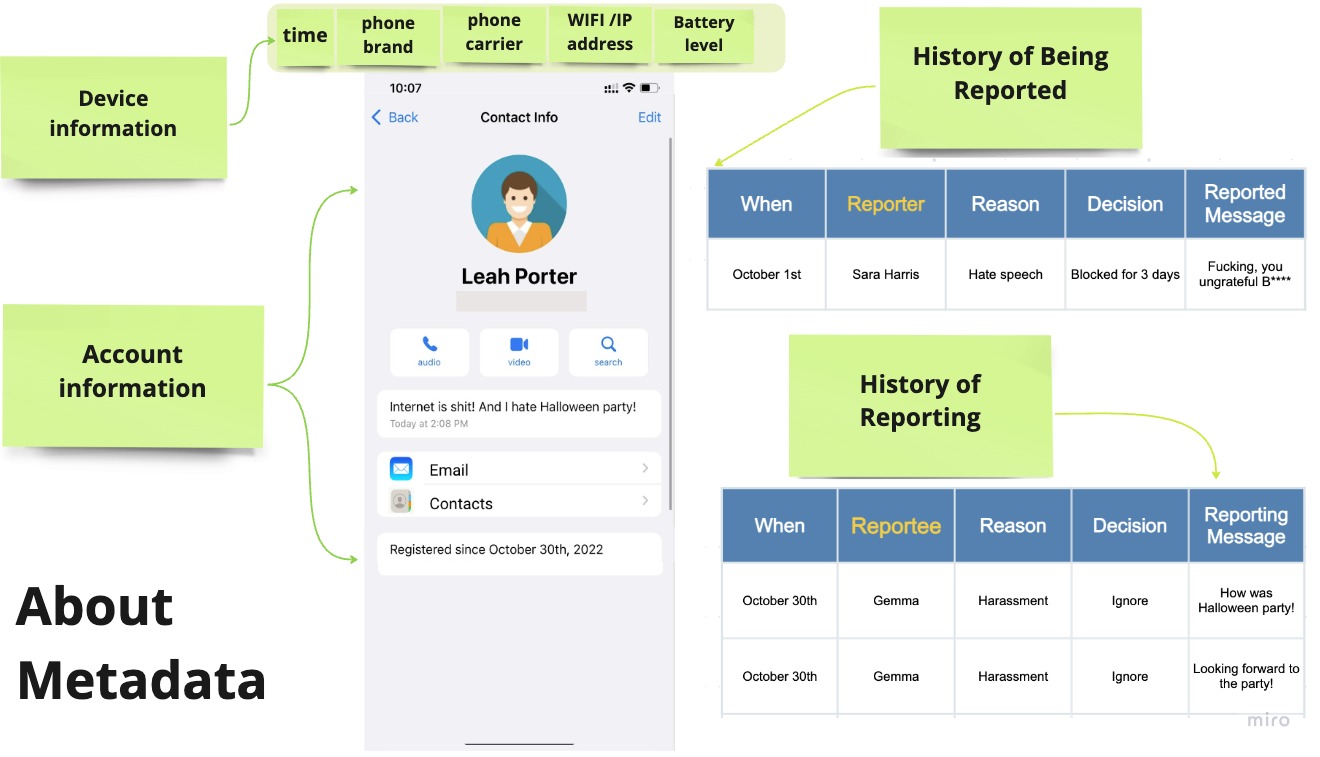}
    \end{subfigure}
    \caption{\textbf{Mock messaging interfaces annotated with cards to illustrate concepts and necessary background information.} The left board introduced to participants different parts of messages that might be shared with E2EE platforms, whereas the right board lists important metadata that might help moderation, including account information (e.g., registration time, phone number, email address), device information, and history of reporting and being reported.}
    \hfill
    \label{onboarding}
\end{figure*}

%% file: sections/03_methodology.tex
\section{Methods}

\subsection{Study design and procedures}
To understand users’ mental models of reporting and their privacy concerns about reporting on E2EE platforms, we conducted semi-structured interviews with active users of E2EE platforms. The final interview protocol was designed iteratively through four pilot interviews to ensure effective elicitation of participants' mental models and contextual concerns. This study was reviewed by our IRB and deemed exempt.

We started the interviews by briefing participants with an overview of the interview session and warning about the possibility of seeing harassment scenarios as part of the interview. We emphasized to participants that they could opt out of questions or stop the interview whenever they wanted, and we gained their explicit consent before we proceeded. We also encouraged participants to think aloud throughout the process. The interview consisted of two sections as described below. The detailed interview protocol can be found in Appendix \ref{appendix}. 

 \textbf{Section I: Mental models of reporting.} In the first section, participants were invited to explain their mental models about how reporting works on E2EE platforms. Inspired by prior work that also investigated mental models \cite{kang2015my, klasnja2009wi, bieringer2022industrial}, we used an interactive card sorting method to better elicit users’ mental models. We created a mock messaging interface based on WhatsApp (Fig. \ref{onboarding}) annotated with cards to help users get familiar with different components relevant to a reporting system. We also created cards labeled with different concepts (the full set of cards are shown in Appendix \ref{appendix}, Fig. \ref{mental models}) on an interactive digital board where users could move around cards to explain their mental models. The digital board prompted participants to reflect on their mental models regarding the following questions: which \textit{stakeholders} have access to reports (e.g., platform, platform moderators, community moderators, hackers, etc.), which \textit{data} is shared (e.g., the reported message, last N messages before the reported message, history of reporting, device information, etc.), and what \textit{moderation action} can be taken (e.g., delete accounts, ban accounts, and delete messages). As participants' perceptions of potential stakeholders are also part of their mental models, we first asked users to explain their understanding of E2EE platforms and to name stakeholders before the card-sorting task.

\begin{figure}[ht]
    \centering
    \includegraphics[width=\columnwidth]{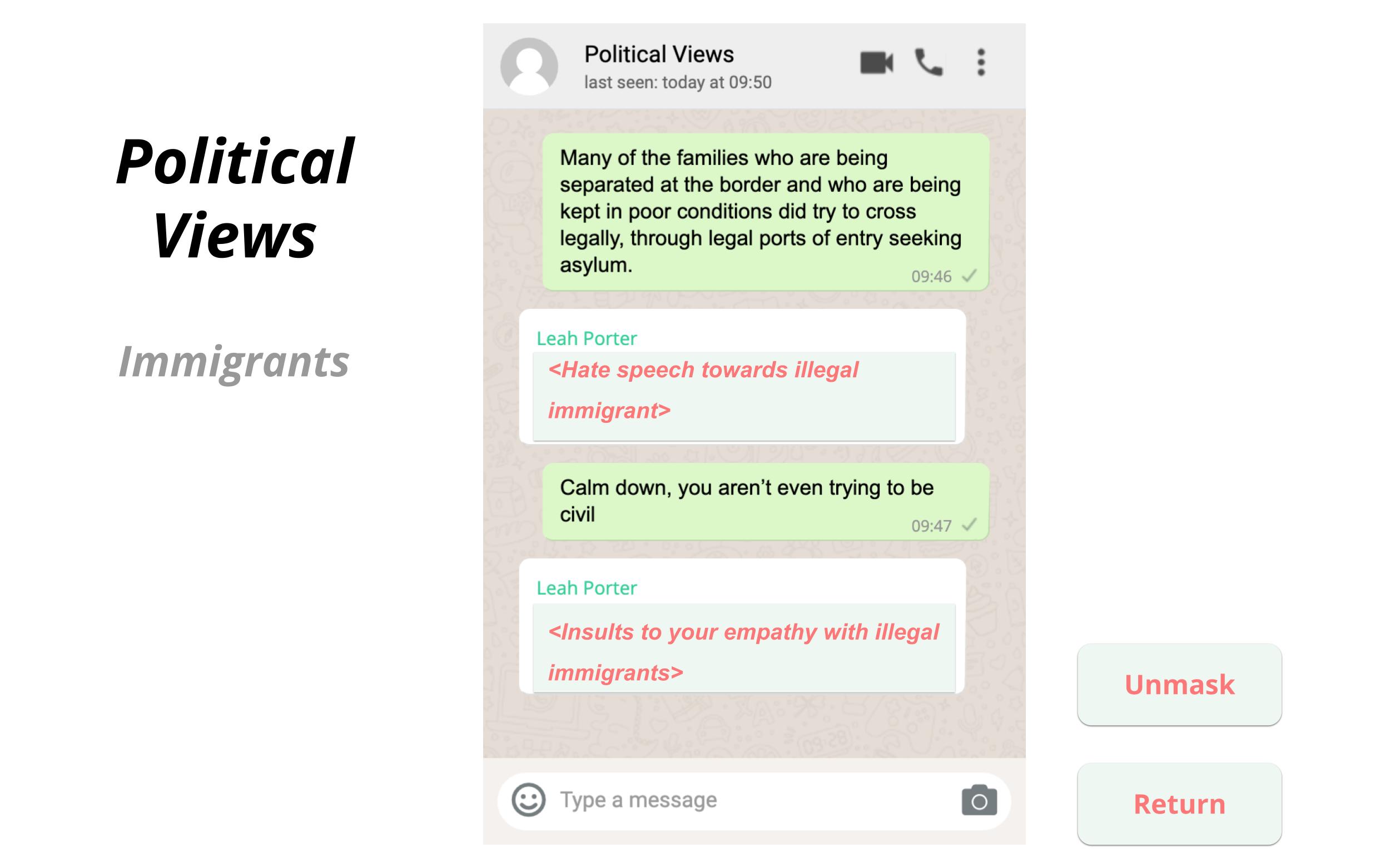}
    \caption{\textbf{Example of a hypothetical harassment scenario.} During the interview, participants were encouraged to select information items that they relate to and are comfortable discussing. Here we present a harassment scenario regarding political views. Note abusive languages are masked by default to protect participants from unnecessary harm.}
    \hfill
    \label{harassment scenarios}
\end{figure}

\begin{table*}[t]
\small
    \centering
    \begin{tabular}{l l l l l l l }
    \toprule
        \textbf{ID} & \textbf{Frequency of}  &\textbf{Reporting} & \textbf{E2EE} & \textbf{Computer} & \textbf{Gender} & \textbf{Race} \\
        & \textbf{Unwanted messages} &\textbf{Experience} & \textbf{Platforms} & \textbf{Literacy} & &
        \\\midrule
        P1 & Every few months & N  & Telegram & Very high & Man & Asian  \\
        P2 & Every few weeks & Y  & WhatsApp & Medium & Woman & Asian  \\
        P3 & About weekly & Y  & WhatsApp, Messenger & Low & Man & White \\
        P4 & Every few months & Y  & Signal, WhatsApp & High & Man & White\\
        P5 & Every few months & Y  & WhatsApp, iMessage & Medium & Man & Asian \\
        P6 & Almost never     & Y  &Signal, WhatsApp & Very high & - & White \\
        P7 & Every few months & N  & Messenger, iMessage & Medium & Man & -\\
        P8 & Almost never & Y  & Signal & Very high & Man & White \\
        P9 & Every few months & Y  & Signal, iMessage & Very high & Woman & White\\ 
        P10 & Almost never & N  & Signal & Medium & - & - \\
        P11 & Almost never & N & Signal & Very high & Man & White \\
        P12 & About weekly & Y & WhatsApp, iMessage & Low & Woman & Asian\\
        P13 & Every few weeks & Y & WhatsApp, Signal & High & Woman & Asian \\
        P14 & Almost never & N & Signal & Very high & Woman & -\\
        P15 & Every few months & Y & iMessage & High & Man & Asian \\
        P16 & Every few weeks & Y & Matrix & Very high & - & - \\
        \bottomrule
    \end{tabular}
     \caption{\textbf{Participant Summary (N=\partiNum)}. A single dash means that the participant preferred not to reveal their demographic information. 
     }
    \label{demographic}
\end{table*}

 \textbf{Section II: Privacy considerations about reporting.} 
As not all of our participants have direct experience with online harassment, we first asked them to consider a series of \textit{hypothetical} harassment scenarios (Fig. \ref{harassment scenarios}) and decide whether and to whom they are going to report, as well as which information they would like to share with moderators. We designed a variety of harassment scenarios that each had a different type of personal information exposed in the context (the full set of scenarios are shown in Appendix \ref{appendix}, Fig. \ref{appendix-harassment scenarios}) so that participants with different experiences can pick scenarios they could better relate to and that they were comfortable viewing \cite{cummings2021need}. 
 Each scenario is structured around the user first revealing some personal information in a message, followed by an exchange where they receive abusive messages related to that information.
 We selected different types of personal information from perceived associated risks found by Milne et al.~\cite{milne2017information}. We then drafted harassment scenarios for each type by first drawing from datasets of hate speech~\cite{vidgen2020directions} and conversation threads on Twitter.
 We further iteratively improved our scenarios via feedback from members of our research lab and pilot interviewees to make them sound more realistic. In order to protect participants from unnecessary exposure to traumatizing content, we masked the abusive texts in each scenario with a high-level description and only unmasked them if participants requested.

\subsection{Recruitment and participants}
We recruited \partiNum{} participants for semi-structured interviews who are active users of E2EE group messaging platforms and preferably have reported unwanted messages on these platforms (See Table \ref{demographic} for detailed demographics). We recruited participants by sharing a recruiting message and a screener survey on Twitter, Mastodon, university-affiliated Slack communities, and privacy-related subreddits including r/europrivacy, r/signal, r/whatsapp, and r/PrivacyGuides.

One challenge of recruitment was that some of our target population are very privacy-aware. To ensure that privacy concerns do not prevent potential participants from signing up for our study, we highlighted our steps to preserve participant privacy in recruiting messages and surveys. For instance, providing demographic information was optional. We only collected participants' email address and name in order to provide compensation, as required by our institution. Participants were also allowed to choose their preferred medium for the interview, including a video call with their camera turned off or via messaging. To observe how participants move cards during the interview, we asked participants to share their screens during the video calls or move cards on Google slides during synchronous interviews over chat. 

From responses to our screener survey, we selected \partiNum{} participants based on their self-described privacy concerns about personal information. We purposefully did not restrict our recruitment to only individuals who have experienced privacy-related online harassment. Instead, we prioritized participants with a diversity of privacy concerns, reporting experiences, and degrees of computer literacy to capture the mental models of a diverse set of people. As a result, we had only a few people who had directly experienced reporting harassment: one participant spoke of a friend who encountered non-consensual imagery, and one participant spoke about receiving political hate speech. 
We discuss the limitations of our participant sample in Section~\ref{sec:limitations}.

We conducted 15 interviews via video calls, where most participants turned their cameras off but had screen sharing on, and 1 via messaging. The interviews lasted 68 minutes on average, and participants were paid \$20. We stopped recruitment when we started hearing repetitive themes and observed no significant new themes.

\subsection{Data analysis}

We analyzed the interview data qualitatively, following the reflexive thematic analysis approach~\cite{Braun2019} to understand participants' mental models of reporting and privacy considerations about reporting. Reflexive thematic analysis has been widely used in HCI research to understand users’ experience, views, as well as factors that influence and shape particular phenomena or processes~\cite{Braun2019}. During data collection, the first author took detailed debrief notes after each interview documenting emerging themes. The authors then collectively reviewed the debrief notes and discussed themes in weekly group meetings. Recordings were automatically transcribed into text. The first author then open-coded the data on a line-by-line basis, and the remaining authors reviewed the transcripts and added codes. Over 350 codes were generated from the open-coding process. The authors clustered the open codes into high-level themes in a codebook and iteratively improved the codebook through discussion. Some examples of codes are \textit{data access of platforms}, \textit{trust in community vs. platform moderators}, and \textit{whether I should report}. Finally, the authors applied the codes to the data to complete the thematic analysis.

%% file: sections/04_findings.tex
\section{Findings}
 We first discuss users' mental models of reporting on E2EE platforms regarding data from user reports. We find that participants expect a limited view of the reported conversation, account information, and device information are shared with platforms (\S~\ref{platform - data access}). Moreover, platform moderators (\S~\ref{platform moderators}) and community moderators (\S~\ref{community moderators}) are expected to only have access to data that are important for reviewing reports. We also find that participants have mixed expectations about how securely the data from user reports might be stored (\S~\ref{data storage}) and used (\S~\ref{data usage}) by E2EE platforms afterward.
 
 Following this, we describe how users make careful reporting decisions to both protect themselves against the privacy risks of online harassment and mitigate the privacy risks of reporting. In particular, we discover that participants believe that reporting may fail to protect their privacy against abusers (\S~\ref{whether report}). Participants also perceive that platform and community moderators play different roles in protecting their privacy, despite having more trust in platform moderators (\S~\ref{to whom report}). Finally, we find that participants are less willing to share personally identifying information but are willing to share more information if they believe reports are anonymized (\S~\ref{which report}).


\subsection{Data Access of Stakeholders}

\subsubsection{Platform} \label{platform - data access}
\textbf{Messages.} Most participants believe that most E2EE messaging platforms have access to the reported message and the contextual messages around it, which they consider to be important for informed moderation decisions. As P3 described, \emph{``the context helps moderators understand what we've been discussing and where the abuse was being perpetrated or where it was taken place.''} Compared to social media platforms in general, E2EE platforms are believed to have more limited access to the context of the reported message. As P11 expressed, \emph{``E2EE platforms get access to that message and the context of that message as well as whatever else they would have to do on my account normally...the contextual might be less on an encrypted message platform.''}

However, some participants also expect E2EE platforms with the most stringent privacy principles to have no access to the context around the reported message or even the content of the reported message. These participants are also more privacy-conscious and opted to use these platforms after deliberation. P14 told us that \emph{``Signal can never see messages unless you send them a screenshot or something. So my preference would be that they have as little access as possible.''} These participants attribute their expectation to the E2EE platform's reputation for privacy protection: \emph{``if somebody finds out that this platform actually shares the last N messages or this entire chat, this press would be too much of a negative thing for the platform [P1].''}

\textbf{Account and conversation information.} Nearly all participants think that, when they report a message, the platform has access to their account information (i.e., their phone number, email address, registration time, and history of reporting and being reported) and relevant conversation information (i.e., who you chatted with, when you chatted, how frequently you chatted with them). But how much account information each E2EE platform collects also varies. For example, P6 believes that platforms with centralized servers can collect more conversation information when relaying messages between ends, while platforms without centralized servers, such as Signal, collect only a little. As P4 said, \emph{``typically they only know when you last logged on and they typically don't even know who sent a message. Of course, they have to know who to deliver the message to. But as I understand, [Signal] knows almost nothing.''} 

 \textbf{Device information.} Most participants expect that most E2EE platforms collect as much device information as possible. P1 explained this more clearly: \emph{``[the platform] might have access to something that don't require extra permissions [such as] battery, model of the phone, advertising, ID. I would expect they would just collect it.''}
 
Almost all participants are not concerned about sharing their account, conversation, and device information (we refer to these as \textit{metadata} in the following analysis) with the platform when reporting unwanted messages for the next two reasons. First, users have low privacy expectations about these kinds of information since most E2EE platforms do not explicitly guarantee the invisibility of metadata to third parties. In fact, most participants believe that platforms have already collected it before any user reporting. P13 expressed such an idea, \emph{``I'm not worried about sharing [metadata] when I report, because I've anyway shared all of this information with them by even using their application. So that information is already with them.''}

 
Second, while concerned about the sensitivity of metadata, some participants acknowledge its importance for reviewing user reports. For example, device information can be used to identify whether the sender uses a fake phone, or is a bot. As P6 expressed, \emph{``device information can be very useful and very identifying. I would like for them not to have it at all probably, but it can help to know if a person is using a fake phone or is using an Android VM and Virtual Box, if it's a bot spamming misinformation messages.''} Similarly, they believe that account information, especially the history of reporting and being reported, is important for reviewing their reports. P5 said: \emph{``So they need account information to sort of build a profile of that person to understand whether they constantly put in fake requests or they constantly report people and that sort of a thing.''}

\subsubsection{Platform moderators}  \label{platform moderators}
Participants have a mixed understanding of the relationship between platform moderators and the platform. Some participants consider these two stakeholders as one entity, thereby believing that platform moderators have the same access as the platform to the information. For example, P2 told us that \emph{``I kind of bunch platform and platform moderators into one entity. I kind of already assume that they have that information because they work for the platform.''}

In contrast, other participants think that platform moderators have a lower level of access to data than the platform. We observe uncertainties among these participants about which subset of data is shared with platform moderators exactly. P5 described his uncertainty as follows, \emph{``To me, the platform knows everything, but I don't have a clear idea about where platform moderators lie in that spectrum of the amount of information that they can access.''} In the following, we discuss information that participants believe platform moderators may have less access to.


\textbf{Account information.}
A considerable number of participants think that platform moderators only have access to non-identifying account information, such as the registration time and history of reporting and being reported, but no personal identifiers like emails, phone numbers, or platform handles. For personal identifiers, participants believe that moderators will \emph{``see some kind of encrypted or hashed kind of version of that number, so they can tell different accounts apart, but they can't kind of reverse engineer who it is [P1].''}

The criteria that participants use to make this distinction is whether they think this piece of account information is important for moderators to make moderation decisions. As P5 has summarized clearly, \emph{``The way I'm thinking about [which information is shared] is what information do I need to make a decision about a specific reported message or a group chat or a person.''} He then gave an example: the history of reporting and being reported can be used to identify false reporters and frequent abusers, but \emph{``other account information isn't as revealing as these two pieces of information.''}
Even for the history of reporting and being reported, several participants expect that a platform that really cares about user privacy would only provide platform moderators with some derivative of this history, such as \emph{``the output of a classification algorithm acting on this information [P14].''}

\textbf{Device information.} Similarly, some participants also expect that only anonymized device information is shared with platform moderators, as evidenced by the words of P11, \emph{``I wouldn't think device information, at least nothing uniquely identifiable, but maybe iPhone or Android or something like that.''} Further, several participants also believe that platform moderators are only provided with some derivative of device information because they are not able to interpret device information on their own. P6 believed that \emph{``The platform would be able to [use device information to] track people across different applications, but the moderators themselves would not be able to use it to perform the tracking.''}

\subsubsection{Community moderators} \label{community moderators}
With a hierarchical structure of reporting systems on messaging platforms in mind, most participants believe that community moderators have access to less data than platform moderators when reviewing user reports. For example, some think that community moderators have access to a limited set of account information, such as the registration time and the location country. They may also only know the history of reports in the scope of their community. P11 implicitly made a distinction between the information community moderators should have versus platform moderators during the interview: \emph{``I'd be fine with a community moderator, for example, knowing how old my account is or maybe what country I'm logged in from, but not something like the email recovery address I use or my IP address. Seems like more of a platform moderator type of information.''}

This distinction is due to participants' perception that community moderators, as active participants in the community, are more familiar with parties involved in the reported conversation than platform moderators. As a result, the platform provides platform moderators with more information to make up for their disadvantages, or equivalently, refrains from giving community moderators unnecessary access to more information. P5 explicitly talked about his perception of community moderators: \emph{``The way I think of community moderators is like a person within the community that's also sending messages and constantly an active participant in that community...They have enough context to just look at the messages and make a decision, rather than platform moderators who need more technical information to kind of detect this kind of thing.''}


\subsection{Data storage} \label{data storage}
Participants expect that the data from user reports will be stored on the platform server for some time. How long the data will be stored depends on local legal requirements, and moderation requirements as moderators need data from previous user reports for future reference. P11 explained his expectations in detail: \emph{``because sometimes the abuse may not be enough to delete the account or ban it so I should be able to act on that information at a later date. I would also imagine for liability reasons they might be required to keep stuff for a certain amount of time.''}

Participants have mixed expectations about the protection of data from user reports by E2EE platforms. Despite acknowledging their limited understanding of technical details, some participants trust E2EE platforms to securely store the data at rest. P13 told us that \emph{``the trust I placed not fully based on understanding technical details, but in trusting them to do the security well.''} 
Other participants, who are more knowledgeable about secure practices, expressed concerns about the platforms' ability to protect the data from potential hacking. For example, P9 was worried that, while the data from reports may be encrypted in transit, it has to be decrypted for moderators' review and might not be encrypted at rest. Platforms may also fail to enforce strict access controls. One participant also expressed concerns that \emph{``even if they had the best intentions and wanted to provide encryption, they might do it for text, but not media, whether it's video or images [P13].''}



\subsection{Data usage} \label{data usage}
In addition to the privacy risks resulting from insecure data storage, participants also believe platforms will further use data from user reports after the reporting procedure for the following purposes. While users do not feel uncomfortable about platforms' developing anti-harassment tools or mining usage patterns based on data from user reports, they are more concerned about being profiled for advertising purposes.

\textbf{Anti-harassment tooling.} Some participants believe that data from user reports will be used to improve anti-harassment tools. Platforms can learn patterns of spurious links in reported messages and block reported accounts that send similar links more proactively. P13 expressed her hopes for this purpose: \emph{``But it's not a worry, it's actually a hope that they would use it towards understanding what issues users face better and trying to make it a safer platform just on the whole.''}

\textbf{Data mining.} Participants also envision that the platform may use the data from user reports to analyze usage patterns and inform its development priorities. For example, the platform can tell which operating systems users are using from device information and prioritize security measures on popular operating systems. The content of messages can reveal cultural interaction patterns as suggested by P14, \emph{``there's a lot of tension between religious groups in India---you can link to the language that they're using, try to see if there's an imbalance that needs to be considered in how you develop your platform strategy.''} Since the data are analyzed in aggregate, participants do not feel uncomfortable about data mining. 

\textbf{Legal investigation.} More than half of the participants believe that data from user reports might be relevant to the investigation of illegal acts (such as terrorism, intimate partner violence, and child abuse) and therefore would be requested by law enforcement agencies. For example, the device and account information of the reported person may help law enforcement agencies identify offenders, and the contextual messages in the reported conversation may also be direct evidence of illegal acts. P14 argued further that the history of user reports could be used to undermine testimony by describing the following example: \emph{``let's say a woman goes to the police and says this guy is stalking me...[by] sending me messages on this platform and waiting outside my house. [But the police] find that you haven't reported any of the messages, suggesting you don't really take it seriously.''}

\textbf{User profiling for targeted ads.} 
Participants have opposing views about whether platforms generate user profiles for ad targeting based on data from user reports. Some participants argue that it is technically and economically infeasible for platforms for two reasons. First, they note that personal information in reported conversations is less organized than the phone number or email address, and therefore platforms need to invest huge computational power, disproportional to the benefits they receive from advertisements, to 
extract personal information. Participants also believe that it is not a comprehensive way to profile users since \emph{``somebody has to either have reported or been reported, and I don't know what percentage of the users are in that group [P10].''}

Other participants believe that it is highly probable for platforms to profile users using data from reports. Platforms can build a granular social graph from conversation information, associate users' metadata with contact information, or even infer their preferences or identities. As P14 suggested, \emph{``if I'm reporting a bunch of homophobia, then the platform could infer that I'm gay with a high probability from that data.''}

Participants also respond with mixed feelings if they know E2EE platforms profit off of generating user profiles from their reports. People who have high privacy expectations about E2EE platforms expressed great disappointment, as P8 clearly stated, \emph{``The way I phrase it to people is your information is valuable, it's financially worth something and they're encouraging you to give them this valuable thing for free.''} In contrast, participants who use E2EE platforms because of peer influence are more indifferent, as P5 explained, \emph{``let's say even if WhatsApp is using that information, I'm already being profiled on so many other platforms that adding in this little thing wouldn't affect my day-to-day life as much.''}

\subsection{Privacy considerations about user reporting on E2EE platforms}

\subsubsection{Whether I should report?} \label{whether report}
We observed that participants compare the privacy risks of reporting with the protection the report can provide to decide whether they should report, as P4 summarized, \emph{``the question is, is reporting this message or this person worth the sacrifice of revealing whatever I've sent.''} For instance, participants tend to report abusive messages if they are in a group chat but tend not to if in a direct message (DM). P2 explained her reasoning in detail: \emph{``For a DM, I don't have to risk myself being identified [by moderators] because of the photo if I don't want to see the photo anymore...I can just delete it on my end and I won't have to see it. But if it's a group chat, then if I delete it on my end, other people can still see it and I don't want other people to still see it. That's why I have to take the risk to report them and have the photo deleted.''}

\textbf{Ineffectiveness.} However, reporting is not always effective in protecting users’ privacy, especially when abusers have already had access to their personal information. For example, when personal photos are used in abusive messages, taking down the photos from the chat cannot prevent further dissemination since abusers can take screenshots or download the photos, and banning the abuser from chat may even motivate them to share photos in more chats or on other platforms. Similarly, when abusers weaponize people's address or phone number to make threats, participants chose not to report to the platform because \emph{``[the most] the platform can do is to probably remove the particular user from the account, but at end of the day from the chats, [the abuser] could already note down my home address [and] compromise my security [P3].''}



\textbf{Inefficiency.} In addition to effectiveness, the efficiency of moderation actions also influences people's decisions to report. When abusers send personal photos of victims in a group chat, P11 explained his consideration to us: \emph{"it's really a question of time ... maybe there's fewer extra people know if I got the message deleted fast enough, but I would have to assume that the community moderators would be very fast to do that.''}





\subsubsection{To whom should I report?} \label{to whom report}
Whether moderators can make an effective and efficient decision to protect users' privacy greatly influences participants' decisions about whom they should report to. Participants think that the following three dimensions contribute to the variance between the effectiveness and efficiency of platform moderators and community moderators. 

\textbf{Moderation areas.} More than half of the participants believe different groups of moderators are responsible for different kinds of user reports. Community moderators are more familiar with \textit{community norms} and therefore can better detect disrespectful behaviors, while platform moderators are more knowledgeable about \textit{platform policies} and can better identify reports of illegal behaviors. P1 explicitly talked about this contrast: \emph{``My mental model of reporting to community moderators is more of a personal issue like [complaints that] `I don't like what's being posted'; my mental model of reporting to platform moderators is more of an illegal stuff like posting child pornography, which violates platform rules.''} 

Therefore, participants tend to report to moderators who they believe can quickly identify the abusiveness of their report and take swift action to protect their privacy. P7 explained his reasoning to us: \emph{``[Abusers] have the potential to do doxxing type activities. So I would want some quick action and it's clearly against stated platform policies. So I would want the platform to take quick action.''}

\textbf{Moderation actions.} Most participants think that platform moderators can delete messages or ban accounts from chat and delete accounts from the platform, while community moderators can do everything but delete accounts from the platform. Hence, participants tend to report to platform moderators if they expect only deleting the abuser's account from the platform can prevent further dissemination of their personal information, as P3 described, \emph{``I believe [platform moderators] are the ones that could solve this problem because I don't need any of these words to be removed and feel the particular account should be deactivated and permanently deleted from the platform.''}


\textbf{Time and resources.} Several participants also believe that community moderators have more time and resources to review user reports than platform moderators. As a result, participants may report to community moderators if they expect more efficient moderation actions to protect their privacy. P6 observed that \emph{``I'll say there are way more community moderators than platform moderators. And also during the time of the night and weekend hours, they would probably be able to moderate while platform moderators are out of duty.''}\\

On the other hand, participants are also concerned that platform or community moderators might abuse the personal information in their reports, such as taking sides with the abuser, leaking information, or doxxing the reporting person. However, a majority of participants have more trust in platform moderators than community moderators to not be abusive or privacy-violating for the following reasons.

\textbf{Expertise.} First, almost all participants perceive platform moderators as more professional in reviewing user reports and trained on how to protect users' privacy. Therefore, platform moderators are believed to recognize the sensitivity of personal information, adhere to codes of conduct about how to access and share information, and refrain from making biased decisions from their personal standpoints. In contrast, as P6 described, \emph{``community moderators are a person voted by the public and probably not versed in privacy and could abuse that information to track down people.''}

\textbf{Accountability.}
Second, many participants think platform moderators are more consistently accountable for potential abuse than community moderators. Platform moderators are supervised by the platform, thus there are more consequences if they violate users' trust; for example, \emph{``getting fired or losing health insurance [P11].''} In contrast, the accountability of community moderators varies across communities. Community moderators who are invested in communities care about their reputation and \emph{``they would have more to lose by losing their reputation and their ability to use the platform [P9],''} while community moderators in a workplace might only be loyal to \emph{``human resources or the management rather than the users and the community itself [P9].''} 

\textbf{Personal connections.}
Moreover, participants also believe that community moderators are more likely to have personal connections with the reporting or the reported person and have motivations to misuse personal information in the reports. P6 gave an example: \emph{``if [community moderators] don't like someone in the chatroom, might do harm with the information with the advantage they possess over some people.''} In comparison, platform moderators are perceived to be distant and unlikely to take a personal grudge against users.

\textbf{Time and resources.}
While having more moderation time and resources can be a motivator for people to report to community moderators, participants also believe that it increases the chances that community moderators might notice the personal information in users' reports and take advantage of it. P4 clearly expressed his concerns: \emph{``I'm assuming that the scope [of community moderators] is smaller and they're responsible for moderating fewer people. My concern would be that they have more time to be worried about me, or they have more time to actually think about this one report.''}


\subsubsection{Which information should I share?} \label{which report}
As we have discussed in \S~\ref{platform - data access}, participants are more concerned about disclosing to moderators the content of their conversations than their account and device information when reporting unwanted messages. In the following, we describe how participants carefully decide which information in their conversation they want to share with moderators to both provide evidence against abusers and protect their own privacy.

\textbf{Towards making an informed moderation decision.} Participants are willing to share information that they believe is important for an informed and fair moderation decision. More than half of the participants chose to share the context around the reported message with moderators to both underline the abusiveness of the reported person and to show their own innocence. For example, P13 suggested that abusive messages can be contextual: \emph{``Sometimes the hate speech can be sarcasm and mockery. It can seem unoffensive when out of context, but with the context, it might make more sense,''} whereas P4 tried to show his civility: \emph{``The other person in this chat is definitely being abusive. But my concern is that if I don't share my own messages, then from the perspective of moderators, I could have also said something horrible.''} 


However, participants tend to share less information and select less sensitive information as long as they believe what they are sharing is enough for an informed and fair moderation decision. For instance, several participants choose not to share the context around the reported message because they believe the reported messages are abusive on the surface and \emph{``the other context might not really help in this position [P3].''} When receiving abusive messages repeatedly, participants choose to only report the abusive messages that do not include their personal information. P2 decided not to report the abusive photo of her: \emph{``Because the whole point of getting this person reported is to have the messages deleted and then banned. And based on the first two [abusive messages without photos], I think it's pretty obvious some action will be taken.''}



\textbf{Less willing to share identifying information.} In general, participants feel less comfortable sharing with moderators messages containing personally identifying information (e.g., phone number, email address, workplace) than those containing information about personal preferences. P4 compared these two kinds of information as follows: \emph{``I think a platform moderator is probably not interested in my personal life. But if I'm revealing where I live, contact information, or my workplace, that would concern me more.''} This is because participants believe identifying information, once put together with information about personal preferences, opens the way for greater abuse, as P10 explained, \emph{``if I make a comment about my political views and it's not attached to any identifier, then you can't trace it back to me and I'm not really concerned about that.''} 

As a result, if participants are certain that their reports are anonymized to platform moderators, they are willing to share more contextual information due to the belief that moderators can not easily associate their personal preferences with their identity. For example, in cases when platform moderators only have access to anonymized reports, P2 told us: \emph{``I'm comfortable sharing the entire chat. But if they have access to my email, to my phone number, to linked social media, then I wouldn't be comfortable sharing all of it.''}

On the other hand, participants are less willing to share information about personal preferences (e.g., medical history, personal photos, political views) with community moderators. In their mental models, participants perceive platform moderators as \emph{``in a far, far away location and removed from the applications infrastructure [P13]''} but community moderators as someone they might run into in their personal life. Therefore, disclosing personal preferences to community moderators not only introduces the risk of abuse but also feels to participants more awkward and confrontational. When receiving abusive messages with his intimate photos, P11 clearly expressed his concerns about context collapse: \emph{``The concern is that when people see the photos you change their impression in some way. [So] I would prefer platform moderators than who I have some relationship with, even if it's in a vague sense that they moderate the community...I'd rather it not be someone who has any ties to me at all.''}

%% file: sections/05_discussion.tex
\section{Discussion}
\subsection{Trust in the reporting system}
In participants' mental models of how reporting works, we observed their uncertainties and privacy concerns, finding that a considerable number of participants feel uncertain about how E2EE platforms will protect data from reports at rest against malicious parties or whether platforms will appropriate data for their unwarranted use. In the end, participants often find themselves \textit{having to} trust these E2EE platforms to act in users' best interest. As P6 summarized concisely, \emph{``No, there is absolutely no guarantee [E2EE platforms] can't get the information. Otherwise, you would compile your own server and own client which nobody would do. So no guarantee, only trusting what the company says and its reputation.''} From the interview, we identified three primary factors that influence people’s trust in E2EE messaging platforms. 

\begin{itemize}[noitemsep,topsep=0pt,leftmargin=*]
    \item \textbf{Open source:} First, participants have more trust in open-source messaging platforms whose source codes and encryption protocols are open to external audit and review. Being open-source also means easy replication and replaceability of the original platform, rendering the platform less likely to violate users' trust. 

    \item \textbf{Business model:} Second, people have more trust in platforms powered by donation and partnership than those powered by advertising and marketing because they believe the latter are more motivated to collect users' data and then generate user profiles~\cite{gerber2018finally}.  

    \item \textbf{Historical behavior:} Finally, historical behavior is another factor that influences people's trust in E2EE platforms~\cite{atwater2015leading, fahl2012helping}. Participants keep an eye on E2EE platforms' behaviors to observe \emph{``a point where they turn and they start acting against their users' interests [P8].''} More privacy-conscious participants further extend their observations to developers and leaders of E2EE platforms. 

\end{itemize}

\noindent Prior research has suggested users' trust in certain technologies may determine whether they can fully utilize the privacy and security advantages these technologies offer~\cite{krombholz2019if, fahl2012helping, naiakshina2016poster}. Here we also found that users' trust in E2EE platforms and moderators significantly influences their reporting decisions. For instance, users make assessments of potential privacy risks of disclosing sensitive information to platforms based on their trust in platforms, which then factor into their decision about whether they should report. We have also observed users' varying levels of trust in platform moderators and community moderators determine to whom they would prefer to report and how much information they would share with them. Some participants may even leave a community if they think moderators are untrustworthy. These findings highlight trust as another important factor in designing a more privacy-preserving reporting system. Prior research has underscored people's mental models of privacy-preserving technologies in shaping their use of these technologies~\cite{abu2018exploring}. However, even if users have a functional mental model of reporting, users may still feel vulnerable because they do not trust the platforms and moderators that operate in this system. Future work should investigate how to increase users' trust in platforms and moderators through design.

\subsection{Privacy calculus in user reporting}

Some of our findings can be analyzed within the broader framework of privacy calculus, a cost-benefit trade-off analysis that accounts for inhibitors and drivers that simultaneously influence the decision on  whether  to  disclose  information or not~\cite{li2010understanding, dinev2006privacy, dinev2006extended}. For instance, our research revealed that individuals' decisions about whether they should report are a result of weighing the privacy risks of reporting (e.g., sharing sensitive information with platforms) against its privacy benefits (e.g., reducing further exposure of sensitive information to more people). Additionally, people make nuanced decisions about whom to report and which information to share in order to minimize the privacy risks of reporting. Similar to prior research in the context of e-commerce transactions, we also observed the impact of trust in platforms on users' evaluation of benefits and risks~\cite{dinev2006privacy}. Future work should further explore other factors, such as individuals' propensity to trust and their control over shared information, within the context of reporting.

\subsection{Design implications for user reporting on E2EE platforms}
\label{design implications discussion}

\begin{figure}
    \centering
    \begin{subfigure}{\columnwidth}
        \centering
        \parbox[c][][c]{\columnwidth}{
            \includegraphics[width=\linewidth]{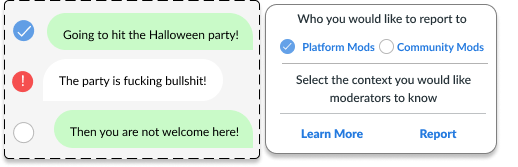}
            \caption{A more granular reporting interface}
        }
        
    \end{subfigure}%
    \\
    \begin{subfigure}{\columnwidth}
        \centering
        \parbox[c][][c]{\columnwidth}{
            \includegraphics[width=\linewidth]{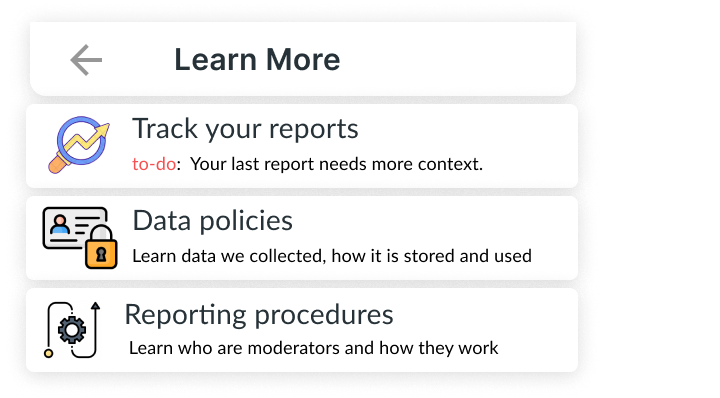}
            \caption{An interface to learn more and interact with moderators}
        }
    \end{subfigure}
    \caption{\textbf{Design implications.} (a) When users choose to report a message, they can select the messages to be included in the report and to whom the report goes. (b) Users can learn more about reporting decisions for individual reports and reporting systems in general, and interact with moderators.}
    \hfill
    \label{design implications}
\end{figure}

Echoing prior research that advocates for providers to focus on users’ needs and experiences when building out their abuse-reporting functionality \cite{kamara2022outside, pfefferkorn2022content}, we further articulate three design implications for user reporting in E2EE platforms and provide a mockup of some of our proposals in Fig. \ref{design implications}. 

 


\subsubsection{More granularity when compiling reports}
Frustrated at their exclusion from the reporting process, nearly all participants desired greater agency when compiling their reports to ``control the contexts in which their information flows'' \cite{marwick2014networked}. First, given the contextual nature of online harassment, participants should be empowered to select which messages to be included in a report. Users could select individual messages in order to exclude messages with sensitive information from a report, or include messages that are important for moderation but outside of the immediate context. We also envision the possibility of obfuscating sensitive information in individual messages. For example, victims of non-consensual intimate imagery might desire to mask their intimate photos but still prove to moderators the existence of these photos. Back-and-forth interactions between users and moderators might also be desirable 
in order to enable \textit{progressive self-disclosure}, allowing users and moderators to reach the right balance between sharing too much and too little. For instance, users could withhold sensitive information on the first try but add additional specific context upon moderators' request. 

Second, users will also benefit from the ability to choose to whom to report, as reporting to different types of moderators has different privacy implications for them. In reality, although users have the choice to report to platform moderators or community moderators, reporting to the latter typically relies on ad hoc approaches such as emails, dedicated channels, or direct messages~\cite{reportDiscord}. Platforms should implement more structured reporting systems to support reporting to community moderators. There is also the possibility that community moderators are malicious or just negligent, and in such cases, the ability to escalate reports to platform moderators after attempting to report to community moderators could help hold community moderators accountable.

However, potential abusers are likely to exploit some of these features to conduct falsified abuse reports~\cite{matias2015reporting}. For example, they may carefully skip the context so that an innocuous joke looks like harassment. While participants consider this possibility as an inevitable sacrifice for more user agency, future work should explore how to uncover falsified abuse reports without compromising users' agency. 

\subsubsection{More interaction between stakeholders} While a report may involve many stakeholders, including community members, the reporting and reported person, and moderators, existing reporting systems only allow a single, one-way interaction between the reporting person and moderators. Future reporting systems should enable more flexible interactions between stakeholders in reporting procedures. For example, the different roles played by platform and community moderators highlight an opportunity for them to interact and collaborate: community moderators can request information such as metadata or cross-community reporting history from platform moderators, who could request information about the context of a report from community moderators in turn. Another idea is to enable the reported person to provide competing evidence and show that the report filed against them is misleading.
Finally, other community members might be asked to corroborate a report, especially in cases where victims of abusive behaviors are too overwhelmed to report all of them~\cite{mahar2018squadbox} or an abusive message is sent in a group chat. These interactions could be one way to gather more information to uncover falsified abuse reports mentioned previously.

However, there are also various risks raised by these interactions, and we emphasize various stakeholders should only be invited after gaining explicit consent from the reporting person~\cite{im2021yes}. 
For instance, the reporting person may not want to have their report shared with platform/community moderators for privacy reasons. Besides, if the reported person is made aware of the report and infers the reporting person, they may choose to retaliate against them. In addition, involving malicious community members in corroborating a report may lead to the report being sabotaged. Therefore, we propose that the reporting person should have the ultimate say in weighing the benefits versus risks and be able to decide whether to invite other stakeholders into the reporting procedure.


\subsubsection{Help develop proper mental models} Our findings indicate that users develop their mental models of reporting and then act accordingly. They may refrain from reporting abusive messages with their sensitive information if they do not believe reports are anonymized to moderators, or from reporting at all if they believe platforms appropriate data from reports for purposes such as advertising. A correct mental model of reporting is also essential if users are provided more agency to make granular reporting decisions, as users may have difficulty understanding the privacy implications of their choices. 

These findings underscore the need to help users develop properly functional mental models. First, platforms should be more transparent about their data policies and procedures of reporting systems. Privacy-conscious participants are eager to know what data is shared by default with the platform versus moderators, and how data is stored and used over time. More disclosure about the workflows of moderation teams and moderation statistics (e.g., how many reports are denied, how frequently each kind of moderation action is made) is also crucial for users to develop trust and confidence in the reporting system. Given users already struggle to understand E2EE itself~\cite{schroder2016signal, abu2017security}, E2EE platforms should provide more tutorials about their reporting systems and use less technical language.


%% file: sections/06_limitations.tex
\section{Limitations and Future Work}
\label{sec:limitations}
To highlight privacy considerations, we narrowed our focus to reporting systems on E2EE platforms instead of online platforms. While future work is needed to determine how these findings generalize to non-E2EE settings, we expect some of our findings to be generalizable. For instance, we saw that participants' decisions about whether to report and to whom to report were not always motivated by their understanding of how E2EE works. This is partly because several participants who use platforms like WhatsApp and Messenger are primarily using them due to peer influence; indeed, some of them had a limited understanding of E2EE \cite{abu2017obstacles, de2016expert}.

There are also limitations regarding our methods. Due to our recruiting method, while we had a few participants with experience with online harassment, most participants only experienced spam on E2EE platforms. While prior research has suggested that privacy-related online harassment is increasingly pervasive~\cite{pewsurveyprivacy, thomas2021sok}, future work should also conduct a more comprehensive survey to understand the landscape of online harassment on E2EE platforms. While we created hypothetical harassment scenarios to help participants put themselves in the shoes of the reporting person, their reporting decisions may still deviate from people who have personally experienced harassment. Future study on new reporting designs should involve more insights and feedback from people who have experienced privacy-related harassment. Given our qualitative approach and purposive sampling, a smaller sample size of \partiNum{} is suitable---however, it also means that the mental models we collected from participants do not cover all E2EE platforms and communities, which may vary greatly in terms of their technical affordances and governance models. Future work may provide a more comprehensive analysis via a larger quantitative study. Moreover, we cannot rule out the possibility that participants’ privacy considerations were implicitly biased by their preconceptions due to our use of a mock interface based on WhatsApp.

Due to the scope of our research, we leave the following research questions for future work. First, while we anecdotally observed that our participants who have experienced reporting harassment could better relate to situations and advocated for greater user agency during interviews, we lacked enough data to compare them with participants without reporting experience to understand how they influenced the study results.
Furthermore, our focus in this research was on examining how individuals' perceived differences between platform and community moderators influence their trust in them. However, given the diverse nature of community settings, future research should investigate the impact of various characteristics of communities on people's trust in moderators.
Finally, as our study focused on gathering insights from community members, we omitted the perspective of community/platform moderators who must decide what action to take based on reports they receive. Due to limited access to platform moderators and opaque internal report handling processes at companies, we intend to address this challenge by interviewing volunteer moderators and server admins on community-operated platforms such as Matrix, Reddit, or Mastodon instead.

%% file: sections/07_conclusion.tex
\section{Conclusions}
Prior research has advocated for user reporting as the moderation approach that most preserves the privacy guarantees of E2EE platforms. However, if users still have privacy concerns or even unfounded misgivings about reporting, user reporting loses its effectiveness in addressing online harassment. Through semi-structured interviews with E2EE users, we uncovered users' mental models and privacy concerns and considerations regarding reporting on E2EE messaging platforms. 
We indeed find that users have privacy concerns about reporting that sometimes lead them to refrain from reporting. Participants also have differing mental models and frequently expressed uncertainty in our interviews about aspects of how reporting works---details that are difficult for the public to validate given the lack of platform transparency. Instead, they often need to rely on their trust in platforms to weigh privacy risks and protections of reporting. Given our findings around the contextual nature of people's privacy concerns, we argue that in order for reporting systems to truly be effective, they need to provide users with a greater ability to navigate trade-offs when it comes to privacy risks.

%% file: sections/08_acknowledge.tex
\section*{Acknowledgments}
This work was supported by the NSF SaTC award \#2120497. We would like to thank Tadayoshi Kohno,  members of the Social Futures Lab at the University of Washington, and collaborators at Cornell Tech for their invaluable help in this project. We also would like to thank our anonymous reviewers for their insightful feedback. Finally, we would like to express our heartfelt thanks to all the participants who dedicated their time and effort to participate in our study.

%% file: sections/09_appendix.tex
\appendix
\section{Appendix}
\label{appendix}
\subsection{Interview Protocol}
Thank you for taking the time to participate in our interview today.  We appreciate your help with this research study. The interview will be about an hour. During the interview, we will ask you about your experience with and thoughts about reporting on encrypted messaging platforms. We will also show you several hypothetical scenarios and ask about your opinions about reporting and related privacy concerns. Please feel free to skip questions or pause the interview if at any point you feel uncomfortable answering the questions.

\medskip

\noindent\textbf{Background: understand the use of E2EE platforms} 

\begin{itemize}[nosep,leftmargin=*]
    \item Why did you start using E2EE messaging platforms? To what extent was privacy protection your motivation for using E2EE platforms? 
    \item Who do you usually chat with on each of these encrypted platforms you use?
    \item Who do you think can have access to your messages on E2EE platforms? 

    \item Have you ever tried to report an account, a message, or a conversation before on these platforms? Do you feel comfortable explaining the context of your reporting? Feel free to skip this question.

    \item Have you ever received or witnessed unwanted messages such as bullying, hate speech, and harassment before on these platforms? Do you feel comfortable describing your experience? Feel free to skip this question.
\end{itemize}

\begin{figure}[h!]
    \centering
    \includegraphics[width=\columnwidth]{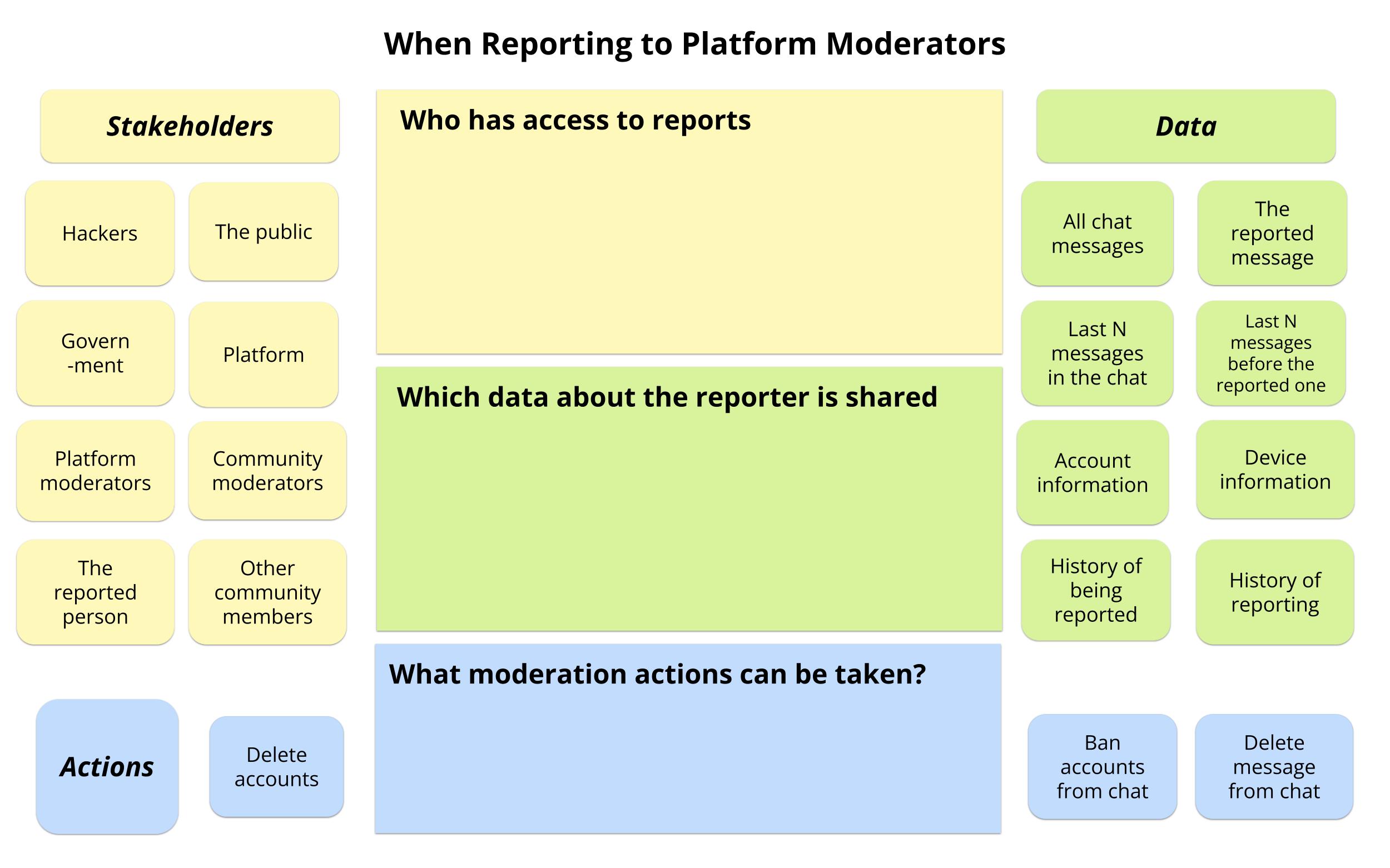}
    \caption{\textbf{Interactive digital board that allows users to explain their mental models of reporting on E2EE.} Participants are encouraged to drag labeled cards to indicate their mental models.}
    \hfill
    \label{mental models}
\end{figure}
\medskip

\noindent \textbf{Section I: eliciting mental models.} For reporting to platform moderators and to community moderators, ask the following questions respectively. During the interview, ask participants to draw the information flow of reports and help them refine the flow by the following interview questions. 

\begin{itemize}[nosep,leftmargin=*]
    \item Who do you think can have access to your reports? 
    \item Which information do you think is shared with moderators when you make a report?
    \item What actions do you think moderators can take for your reports?
    \item Are you worried about these data will be shared with each stakeholder here? 
    \item Do you imagine that these shared data will still be used elsewhere after report decisions are made?
    \item Have you noticed or experienced falsified abuse reports?
\end{itemize}

\medskip
\begin{figure}
    \centering
    \includegraphics[width=\columnwidth]{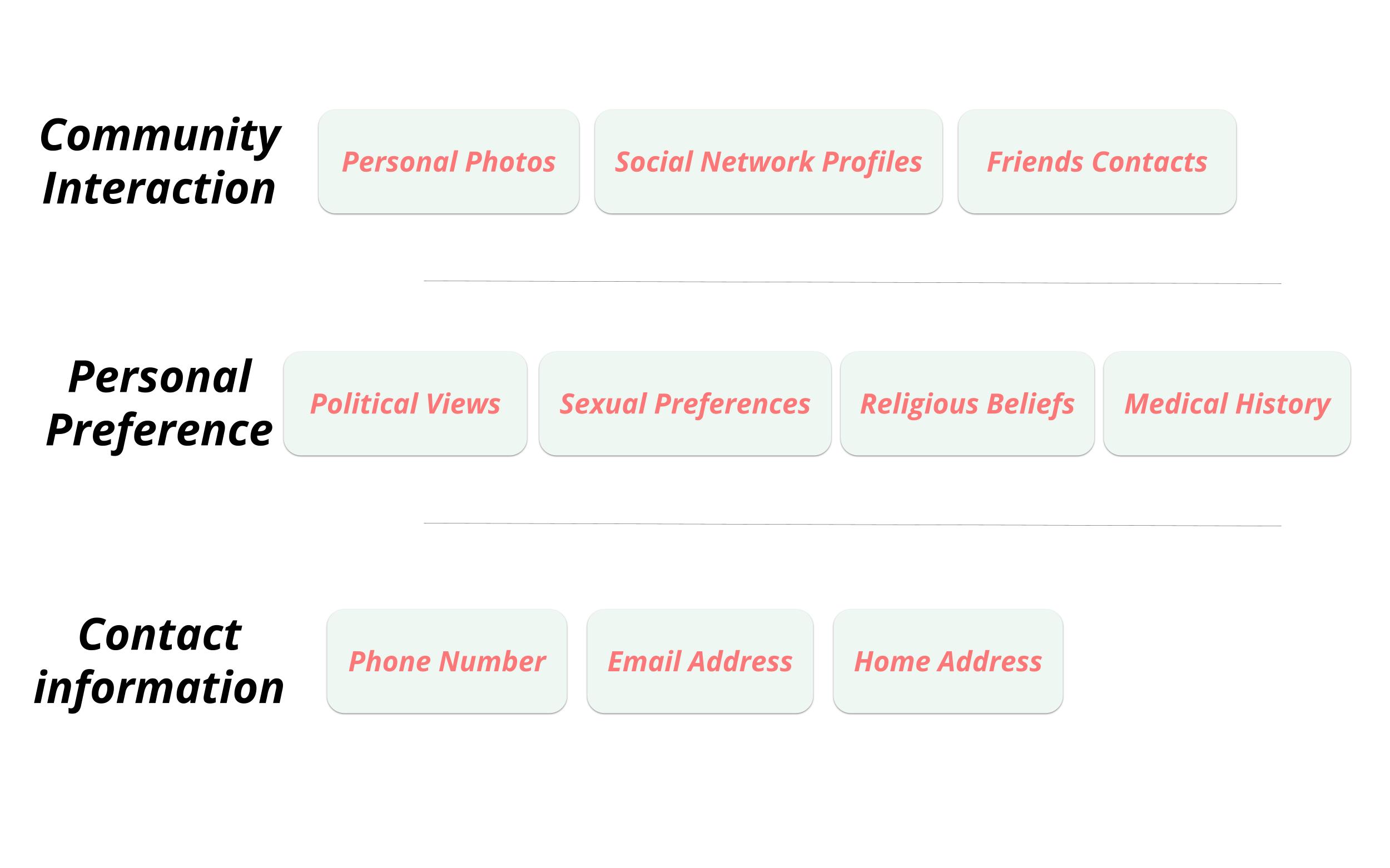}
    \caption{\textbf{Harassment scenario options.} During the interview, participants were encouraged to select the items that they relate to and are comfortable discussing from this board listing all the possible options in a hypothetical harassment scenario.  Abusive language is masked by default to protect participants from unnecessary harm.}
    \hfill
    \label{appendix-harassment scenarios}
\end{figure}

\noindent \textbf{Section II: ask privacy concerns about reporting.} First, ask participants to choose the category that they share frequently and consider sensitive but also feel comfortable talking about. In each selected scenario, ask the following questions with the introduction: Imagine you are using an E2EE group messaging platform like WhatsApp or Signal. You received a few abusive messages from another chat member. 

\begin{itemize}[nosep,leftmargin=*]
    \item Will you choose to report these messages or not?
    \item Who are you going to report to, community moderators or platform moderators?

    \item What actions do you think moderators can take for your reports?
    \item If you are allowed to choose which part of messages is filed in a report, what messages would you like to choose for a report?
    \item If you are allowed to choose which part of metadata is filed in a report, which metadata information about you and the reported person do you would like to choose for a report?
\end{itemize}

\subsection{Hypothetical Harassment Scenarios}
\textit{\color{red} \textbf{Content Warning:} The following scenarios contain hate speech and offensive language.}

\medskip

\noindent\textbf{Political views}\\
A: Many of the families who are being separated at the border and who are being kept in poor conditions did try to cross legally, through legal ports of entry seeking asylum.

\noindent B: They are their own country's problems. Not the problems of the United States. They are fucking invaders. [description: \textit{B expressed hate speech about illegal immigrants}]

\noindent A: Calm down, you aren’t even trying to be civil.

\noindent B: Then deport you big trash and your friends to Mexico. If you want to help the immigration problem, start there! Problem solved for all, you pussy liberals [description: \textit{B expressed hate speech about people’s empathy towards illegal immigrants}]

\medskip 

\noindent\textbf{Sexual preference}\\
A: Hey everyone, I just want to share that I am a woman and go by Samantha and use she/her pronouns.

\noindent B: Ugh, that’s fucking disgusting, you are a boy, you have a penis, and you are Robert. you cannot just change your name and gender. [description: \textit{B: Insults to you about your being transgender}]

\noindent A: I had hoped that you’d be more supportive.

\noindent B: This is so sad and pathetic, you are such a loser and a sissy, you are dead to me. And I will tell everyone about this. [description: \textit{B expressed insults to you about your being transgender and threatens to tell everyone}]

\medskip 

\noindent\textbf{Religious belief}\\
A: As a Muslim girl, am I welcome here?!

\noindent B: No, you are not! We did NOT invite Muslims, Africans, and all sorts here! Fuck that Muslim piece of shit. [description: \textit{B expressed hate speech towards Muslim people}]

\noindent A: I don't believe in violence and we should treat each other with love.

\noindent B: As Muslims, go back to your Muslim hell hole countries, you ungrateful B****!  [description: \textit{B expressed insults to you about your being Muslims}]

\medskip 

\noindent\textbf{Personal photos}\\
A: Just getting some cake from the Cafe down the street. (a photo of yourself holding a cake)

\noindent B: You are so fat, why do you eat so much? [description: \textit{B expressed insults to you about your body}]

\noindent A: Excuse me?

\noindent B: I am going to re-share this edited photo to Facebook (a photo of A holding a cake; now with the caption “fat cat, fatty cat”)  [description: \textit{B expressed insults to you about your body and shared an edited but now abusive photo of you holding the cake}]

\medskip 

\noindent\textbf{Phone number}\\
(Context: A shared their phone number with B when they were on a company vacation together, the internet was spotty, and they needed to use SMS to communicate.)

\smallskip

\noindent A: I am planning on visiting my family in Turkey during this holiday, and may be unavailable for the next few days.

\noindent B: Ugh, I didn’t know that you were an immigrant. Immigrants are smelly, shitty, and taking over our jobs. [description: \textit{B expressed hate speech towards immigrants}]

\noindent A: Wait, what?

\noindent B: Get back to your country. I am going to sign up for Tinder with your phone number +1 2045661223 and swipe right on every guy. [description: \textit{B expressed hate speech towards immigrants and threats to overwhelm you with spam and stalkers using your phone number}]

\medskip 

\noindent\textbf{Email address}\\
(Context: A shared their work email address with B for an office meeting.)

\smallskip

\noindent A: I am planning on visiting my family in Turkey during this holiday, and may be unavailable for the next few days.

\noindent B: Ugh, I didn’t know that you were an immigrant. Immigrants are smelly, shitty, and taking over our jobs. [description: \textit{B expressed hate speech towards immigrants}]

\noindent A: Wait, what?

\noindent B: Get back to your country. I am going to sign up for gay twink porn with your work email address alice@bigcompany.co  [description: \textit{B expressed hate speech towards immigrants and threats to overwhelm you with spam and stalkers using your email address}]

\medskip 

\noindent\textbf{Home address}\\
(Context: A shared their home address with B for the office secret Santa list.)

\smallskip

\noindent A: I am planning on visiting my family in Turkey during this holiday, and may be unavailable for the next few days.

\noindent B: Ugh, I didn’t know that you were an immigrant. Immigrants are smelly, shitty, and taking over our jobs. [description: \textit{B expressed hate speech towards immigrants}]

\noindent A: Wait, what?

\noindent B: Get back to your country. I know that your address is 4200 11th Ave NE, XXX. I am going to call the police to your address to forcefully kick you out. [description: \textit{B expressed hate speech towards immigrants and threats to overwhelm you with false police calls using your home address}]